\documentclass{cicirello}
\pdfoutput=1

\usepackage[numbers]{natbib}
\usepackage{doi}
\usepackage{listings}
\usepackage[noend]{algorithmic}
\usepackage{algorithm}
\title{Fast Gaussian Distributed Pseudorandom Number Generation in Java via the Ziggurat Algorithm}
\author{Vincent A. Cicirello}
\orcid{0000-0003-1072-8559}
\reportnum{ALG-24-009}
\reporturl{https://reports.cicirello.org/24/008/ALG-24-009.pdf}
\copyrightyear{2024}
\date{May 2024}

\address{Computer Science\\
Stockton University\\
Galloway, NJ 08205 USA\\
\url{https://www.cicirello.org/}
}

\ACM{F.2.1; I.1.2; D.2.13; G.3; G.4}

\abstract{We report on experiments with the ziggurat algorithm
for generating Gaussian distributed random numbers. The study
utilizes our open source Java implementation that was introduced 
originally for Java 11 at a time when the Java API only provided 
the much slower polar method. Our Java implementation of the ziggurat
algorithm is a port of the GNU Scientific Library's C implementation.
Java 17 introduced a significant overhaul of pseudorandom number
generation, including several modern pseudorandom number generators (PRNGs)
as well as additional functionality, among which includes switching
from the polar method to a modified ziggurat algorithm. In the experiments 
of this paper, we explore whether there is still a need for our implementation 
for Java 17+ applications. Our results show that Java 17's modified ziggurat 
is faster than our implementation for the PRNGs that support it. However, 
Java 17+ continues to use the polar method for the legacy PRNGs Random, 
SecureRandom, and ThreadLocalRandom. The linear congruential method of 
Java's Random class lacks the statistical properties required by Java's 
modified ziggurat implementation; and SecureRandom and ThreadLocalRandom 
unfortunately use the polar method as a side-effect of extending Random. Our 
implementation of the original ziggurat algorithm does not require the same 
statistical properties of the underlying PRNG as Java 17's optimized version, 
and can be used with any of these PRNGs, and is especially relevant where 
pre-Java 17 support is required.}

\begin{document}

\maketitle

\section{Introduction}

Efficiently generating Gaussian distributed random numbers is often important
in modeling and simulation contexts. The most commonly encountered algorithm
for generating Gaussian random variates is the polar method~\cite{Knuth2}, 
which has been around for decades. However, there are now modern alternatives
that are significantly faster and provide higher quality results, such as the
ziggurat method and its 
variations~\cite{Marsaglia2000,Leong2005,Voss2014,McFarland2014}.

Our original motivation for efficiently generating Gaussian random variates
was in the context of implementing Gaussian mutation~\cite{Hinterding1995} for
an evolutionary algorithm (EA)~\cite{cicirello2018flairs}. The EA in question was implemented
in Java, which at the time (e.g., Java 8) provided only the polar method
for Gaussian random number generation. Runtime profiling of the EA revealed
this as a bottleneck. We ported the GNU Scientific Library's C 
implementation~\cite{Voss2014} of the ziggurat algorithm~\cite{Marsaglia2000,Leong2005}
to Java to optimize the EA runtime. We also carefully considered other
random number related elements of the EA, such as the choice of pseudorandom random
number generator (PRNG) itself. We demonstrated that an EA relies so heavily on
random number generation that switching to the ziggurat algorithm and making
a few other PRNG related optimizations resulted in an EA that is 20\% to 25\% 
faster~\cite{cicirello2018flairs}. We since released our Java implementation of
the ziggurat algorithm as a small open source library, which we utilize in the
experiments of this report. See Table~\ref{tab:zig} for 
URLs to the source code as well as to the artifacts in the Maven Central Repository.

\begin{table}[t]
\caption{Important URLs for the open source ziggurat library}\label{tab:zig}
\centering
\begin{tabular}{ll}\hline
Source & \url{https://github.com/cicirello/ZigguratGaussian} \\
Maven & \url{https://central.sonatype.com/artifact/org.cicirello/ziggurat} \\
DOI & \url{https://doi.org/10.5281/zenodo.4106912} \\\hline
\end{tabular}
\end{table}

Among other things, Java 17 overhauled random number generation~\cite{Steele2017}, 
including introducing several modern PRNGs~\cite{Blackman2021,Steele2021}, a hierarchy 
of interfaces for different categories of PRNG, as well as other random number improvements.
One of those improvements was providing an implementation of McFarland's modified ziggurat
algorithm~\cite{McFarland2014} for Gaussian random variates that is faster than 
the original. It did not totally replace the polar method, however, as the \lstinline|Random|
class's linear congruential generator (LCG)~\cite{Knuth2} does not meet the quality of
randomness required by the modified ziggurat method. Thus, \lstinline|Random| retains
the slow polar method. The other legacy PRNG classes that extend \lstinline|Random| have
unfortunately also retained the polar method, as a consequence of subclassing \lstinline|Random|,
despite providing the required quality of randomness.

In this paper, we report on experiments comparing our implementation of the original ziggurat
algorithm with the Gaussian random variate implementations provided in Java 17, which is the
polar method for some PRNG classes, and the modified ziggurat for others. Unlike
McFarland's modified ziggurat, our implementation of the original ziggurat is applicable for
all of Java's PRNG classes, as it does not rely on the low-order bits of random 64-bit longs.
We will see that Java's new modified ziggurat is faster than the original ziggurat in cases 
where it is applicable. But in cases where the modified ziggurat is not applicable
(e.g., when used with a LCG) or where the modified ziggurat is not otherwise provided that
the original ziggurat as implemented in our library is significantly faster than the alternative,
and continues to be relevant for applications that must support pre-Java 17 runtime environments.

We proceed as follows. We explain our methodology in Section~\ref{sec:methodology}. 
We discuss the experimental results in Section~\ref{sec:results}.
We conclude in Section~\ref{sec:conclusion} with a discussion and observations of
when our ziggurat library is still applicable, and the better alternatives
that are available in other cases.

\section{Methodology}\label{sec:methodology}

We use OpenJDK 64-Bit Server VM version 17.0.2 in the experiments on a Windows 10 PC
with an AMD A10-5700, 3.4 GHz processor and 8GB memory. We experiment with three of 
Java's legacy PRNGs:
\begin{itemize}
\item \lstinline|Random|: implementation of a LCG~\cite{Knuth2};
\item \lstinline|SplittableRandom|: implementation of SplitMix~\cite{Steele2014} algorithm, 
which is a faster optimized version of the DotMix~\cite{Leiserson2012} algorithm, and which passes the 
DieHarder~\cite{brown2013dieharder} tests; and
\item \lstinline|ThreadLocalRandom|: implementation of the same PRNG algorithm 
as \lstinline|SplittableRandom|, but manages thread local seed data in multi-threaded
scenarios.
\end{itemize}
For Gaussian random number generation, and for each of the above PRNGs, 
we consider the following:
\begin{itemize}
\item Ziggurat: This is our implementation (see Table~\ref{tab:zig}) of the ziggurat 
algorithm~\cite{Marsaglia2000,Leong2005}, which is a Java port of the GNU Scientific
Library's C implementation~\cite{Voss2014}.
\item Java 17: We compare our ziggurat implementation to Java 17's builtin support for
Gaussian random numbers, which is the polar method~\cite{Knuth2} for the \lstinline|Random|
and \lstinline|ThreadLocalRandom| classes, and is McFarland's modified ziggurat 
algorithm~\cite{McFarland2014} for \lstinline|SplittableRandom|.
\item Wrapped: For \lstinline|ThreadLocalRandom|, we consider a third option where we coerce the
use of Java 17's modified ziggurat implementation instead of the polar method by 
implementing the \lstinline|RandomGenerator| interface in a way that wraps calls to 
\lstinline|ThreadLocalRandom.current().nextLong()| in the only required method of the 
\lstinline|RandomGenerator| interface, the \lstinline|RandomGenerator.nextLong()| method.
This forces the use of the default \lstinline|RandomGenerator.nextGaussian()| method, which
is McFarland's modified ziggurat algorithm~\cite{McFarland2014}, but utilizing
\lstinline|ThreadLocalRandom| as the PRNG. See Listing~\ref{lst:wrapped}.
\end{itemize}
Listing~\ref{lst:calls} illustrates usage of our ziggurat implementation versus the
Java API's builtin Gaussian support.

\begin{lstlisting}[float,frame=lines,caption={Coercing \lstinline|ThreadLocalRandom| to use Java 17's modified ziggurat algorithm},label={lst:wrapped}]
RandomGenerator wrappedThreadLocalRandom =
  new RandomGenerator() {
    @Override
    public long nextLong() {
      return ThreadLocalRandom.current().nextLong();
    }
  };

// You can then generate Gaussian random numbers using Java 17's 
// modified ziggurat with ThreadLocalRandom, with calls like:
double r1 = wrappedThreadLocalRandom.nextGaussian();

// Whereas calls like the following would use the polar method:
double r2 = ThreadLocalRandom.current().nextGaussian();
\end{lstlisting}

\begin{lstlisting}[float,frame=lines,caption={Using our ziggurat implementation vs Java's builtin functionality},label={lst:calls}]
Random random = new Random();
SecureRandom secure = new SecureRandom();
SplittableRandom splittable = new SplittableRandom();

// Calls to nextGaussian() for Random, SecureRandom, and ThreadLocalRandom 
// classes use the polar method:
double g1 = random.nextGaussian();
double g2 = secure.nextGaussian();
double g3 = ThreadLocalRandom.current().nextGaussian();

// The equivalent call for SplittableRandom uses Java's modified ziggurat:
double g4 = splittable.nextGaussian();

// To use our implementation of the original ziggurat algorithm with Random, SecureRandom,
// ThreadLocalRandom, and SplittableRandom, respectively, pass the PRNG instance to
// ZigguratGaussian.nextGaussian as a parameter, or no parameter for ThreadLocalRandom:
double z1 = ZigguratGaussian.nextGaussian(random);
double z2 = ZigguratGaussian.nextGaussian(secure);
double z3 = ZigguratGaussian.nextGaussian();
double z4 = ZigguratGaussian.nextGaussian(splittable);
\end{lstlisting}

We use the Java Microbenchmark Harness (JMH)~\cite{JMH137} to implement our experiments. 
For each experiment condition (combination of PRNG and Gaussian implementation), we use 
five 10-second warmup iterations to ensure that the 
Java JVM is properly warmed up, and we likewise use five 10-second iterations for measurement.
We measure and report average time per operation in nanoseconds, along with 99.9\% confidence 
intervals.

The code to reproduce our experiments is available on GitHub, as is the data from our
runs of the experiments. See Table~\ref{tab:reproducible} for the relevant URLs.

\begin{table}[t]
\caption{Reproducible results: URLs to experiment code and data}\label{tab:reproducible}
\centering
\addtolength{\tabcolsep}{-1pt}
\begin{tabular}{ll}\hline
Code & \url{https://github.com/cicirello/ZigguratGaussian/experiment/timing17} \\
Data & The file \href{https://github.com/cicirello/ZigguratGaussian/experiment/timing17/results17.txt}{/results17.txt} in above directory \\ \hline
\end{tabular}
\addtolength{\tabcolsep}{1pt}
\end{table}

\section{Results}\label{sec:results}

Table~\ref{tab:results} summarizes the results. Our implementation of the ziggurat algorithm for
generating Gaussian distributed random numbers is 83.12\% faster than Java 17's polar method implementation
when the legacy \lstinline|Random| class is used; and our ziggurat implementation is 87.72\% faster than
Java 17 in the case of the \lstinline|ThreadLocalRandom| class. We expect that we would find approximately
the same significant advantage in the case of Java's \lstinline|SecureRandom| class, although we did not test
this, since that class likewise uses the slow polar method.

\begin{table}[t]
\caption{Results: Average time per operation in nanoseconds}\label{tab:results}
\centering
\begin{tabular}{lrrr}\hline
PRNG class                    & Ziggurat           & Java 17                         & Wrapped \\ \hline
\lstinline|Random|            & $17.393 \pm 0.300$ ns/op & $103.037 \pm 0.951$ ns/op & n/a \\
\lstinline|SplittableRandom|  & $10.089 \pm 0.139$ ns/op & $8.927 \pm 0.026$ ns/op   & n/a \\
\lstinline|ThreadLocalRandom| & $10.901 \pm 0.032$ ns/op & $88.774 \pm 2.096$ ns/op  & $10.392 \pm 0.121$ ns/op \\
\hline
\end{tabular}
\end{table}

The case of \lstinline|SplittableRandom|, however, is different. In Java 17, the
\lstinline|SplittableRandom| class uses McFarland's modified ziggurat 
algorithm~\cite{McFarland2014}, which in our experiments is 11.52\% faster than
our implementation of the older version of the ziggurat algorithm. When we use
our trick from Listing~\ref{lst:wrapped} to coerce the use of Java 17's modified
ziggurat algorithm for the \lstinline|ThreadLocalRandom| class, we find that it is
4.67\% faster than our implementation of the original ziggurat algorithm. Java 17
introduced several modern PRNGs~\cite{Steele2017}, including Xor-Based Generators 
(XBG)~\cite{Blackman2021} and several variations of LXM~\cite{Steele2021}, which
combine an LCG and XBG with a mixing function. All of these newly introduced PRNGs
use Java 17's modified ziggurat algorithm when generating Gaussian random numbers.
So although we did not include these in our experiments, we should find the same
pattern as we found with \lstinline|SplittableRandom|, namely that Java's modified
ziggurat should be a bit faster than our implementation of the original ziggurat
algorithm.

Java 17's implementation of McFarland's modified ziggurat depends on the quality of
the low-order bits from calls to the \lstinline|nextLong()| method. The
\lstinline|ThreadLocalRandom| class meets those expectations, as it implements the
same PRNG algorithm as the \lstinline|SplittableRandom| class. Its use of the slow
polar method is due to subclassing the \lstinline|Random| class, whose use of an
LCG leads to low-order bits that do not meet the requirements
of Java's modified ziggurat implementation. Thus, it is safe to use our trick from
Listing~\ref{lst:wrapped} with the \lstinline|ThreadLocalRandom| class, but not with
the \lstinline|Random| class. Our implementation of the original ziggurat algorithm, 
as ported to Java from the C implementation of the GNU Scientific Library, does not 
depend on the quality of low-order bits of random longs and should be safe to use 
with any of the PRNG classes.

\section{Conclusions}\label{sec:conclusion}

From our experimental results, we make the following observations:
\begin{itemize}
\item \begin{sloppypar}Our open source implementation of the original ziggurat algorithm (see library details 
in Table~\ref{tab:zig}) provides a very significant performance advantage for the 
\lstinline|Random|, \lstinline|ThreadLocalRandom|, and \lstinline|SecureRandom| classes as 
compared to the Java API's polar method implementation, with our results (Table~\ref{tab:results}) 
showing that our ziggurat implementation is over 83\% and 87\% faster in the cases of 
\lstinline|Random| and \lstinline|ThreadLocalRandom|, respectively. Unlike Java 17's modified 
ziggurat, our ziggurat implementation is applicable even in the case of weaker PRNGs like the 
LCG implemented by \lstinline|Random|.\end{sloppypar}

\item \begin{sloppypar}For pre-Java 17 applications, our ziggurat library provides a Gaussian 
implementation for \lstinline|SplittableRandom|, where the Java API itself lacks Gaussian support 
entirely.\end{sloppypar}

\item For Java 17+ applications, Java's modified ziggurat algorithm, implemented by the
default interface method \lstinline|RandomGenerator.nextGaussian()| is more than 11\% faster 
than our implementation of the original ziggurat in the case of the \lstinline|SplittableRandom|
class (Table~\ref{tab:results}), and presumably for all of the modern 
PRNGs~\cite{Steele2017,Steele2021,Blackman2021} introduced in Java 17. Note that we did not
verify this for the new Java 17+ PRNGs as our ziggurat library is designed to support Java 11+,
which does not have the \lstinline|RandomGenerator| interface. Thus, modifying our ziggurat
library to support the \lstinline|RandomGenerator| interface would break Java 11 compatibility.

\item For Java 17+ applications that use \lstinline|ThreadLocalRandom|, another faster option
is to utilize our trick from Listing~\ref{lst:wrapped} to gain access to Java's implementation
of the modified ziggurat algorithm. Although in such a case, utilizing our ziggurat library may
be a cleaner approach while almost as fast.
\end{itemize}

A more comprehensive library of randomization utlities $\rho\mu$~\cite{cicirello2022joss}
has used our implementation of the ziggurat algorithm for $\rho\mu$ versions 1.0.0 through 
4.0.0. However, beginning with version 4.1.0, the $\rho\mu$ library has been updated based on the
insights of these experiments. It now relies on the Java API's modified ziggurat by default
for all PRNGs that support it, and for \lstinline|ThreadLocalRandom| it uses the
Listing~\ref{lst:wrapped} trick as a replacement for our ziggurat implementation. It does, however,
maintain a utility class with our implementation of the original version of the ziggurat algorithm
to continue to support fast Gaussian random number generation for the legacy \lstinline|Random|
and \lstinline|SecureRandom| classes. The $\rho\mu$ library requires a minimum version of 
Java 17, however, so applications supporting pre-Java 17 may still benefit from our ziggurat
library, which only requires Java 11+. See Table~\ref{tab:rhomu} for relevant URLs to the
$\rho\mu$ library. Among other purposes, $\rho\mu$ supports our original motivation of
accelerating the runtime of EAs by optimizing random number generation, serving as a dependency
of the open source evolutionary computation library Chips-n-Salsa~\cite{cicirello2020joss}.

\begin{table}[t]
\caption{Important URLs for the open source $\rho\mu$ library}\label{tab:rhomu}
\centering
\begin{tabular}{ll}\hline
Source & \url{https://github.com/cicirello/rho-mu} \\
Maven & \url{https://central.sonatype.com/artifact/org.cicirello/rho-mu} \\
Website & \url{https://rho-mu.cicirello.org/} \\\hline
\end{tabular}
\end{table}

\bibliographystyle{plainnat}
\bibliography{zig}
\end{document}